\documentclass[11pt]{article}
\usepackage[T1]{fontenc}
\usepackage[utf8]{inputenc}
\usepackage[bottom=4cm, headsep=1cm, footskip=1.75cm]{geometry}

\usepackage[colorinlistoftodos]{todonotes}

\usepackage{tabularx}
\usepackage{makecell, array, colortbl, booktabs, multirow}
\newcolumntype{L}[1]{>{\raggedright\let\newline\\\arraybackslash\hspace{0pt}}m{#1}}
\newcolumntype{C}[1]{>{\centering\let\newline\\\arraybackslash\hspace{0pt}}m{#1}}
\newcolumntype{R}[1]{>{\raggedleft\let\newline\\\arraybackslash\hspace{0pt}}m{#1}}

\usepackage[format=plain, textfont=it, labelfont={it, bf}]{caption}

\usepackage{graphicx, subcaption}

\usepackage[sectionbib]{natbib}
\usepackage{chapterbib}

\usepackage{amsmath, amsfonts, amssymb, amsthm, nicefrac, bbm, bm, stmaryrd}
\theoremstyle{theorem}

\theoremstyle{remark}

\usepackage{algorithm, algpseudocode}

\usepackage{soul}

\usepackage{xcolor}
\definecolor{blu1}{HTML}{4285f4}
\definecolor{rosso1}{HTML}{ea4335}
\definecolor{giallo1}{HTML}{fbbc04}

\definecolor{col1}{HTML}{6118b5}
\definecolor{col2}{HTML}{f09ec2}

\usepackage{lscape, afterpage}

\usepackage{setspace}
\usepackage{changepage}

\PassOptionsToPackage{hyphens}{url}\usepackage{hyperref}
\hypersetup{
	colorlinks,
	citecolor=blu1,
	filecolor=black,
	linkcolor=black,
	urlcolor=blu1
}

\usepackage{tikz, pgfplots}
\pgfplotsset{compat = 1.16}
\usepgflibrary{shapes.arrows}
\usepgfplotslibrary{statistics}
\usetikzlibrary{
	tikzmark,
	arrows,
	calc,
	decorations.pathreplacing,
	calligraphy,
	matrix,
	positioning,
	intersections,
	pgfplots.fillbetween,
	shadows,
	automata,
	decorations.markings,
	petri,
	shapes.geometric,
	fadings
}

\usepackage{xfp}
\newlength\WIDTHOFBAR	

\usepackage{titlesec, xhfill}
\definecolor{verdescuro}{HTML}{001c08}
\colorlet{rulecolor}{verdescuro}

\def\bl#1{\mbox{\tiny $\bm{#1}$}} 

\title{Dealing with separation problem in hidden Markov models with covariates based on a  penalized maximum likelihood approach}

\author{Luca Brusa$^1$, Fulvia Pennoni$^1$, Francesco Bartolucci$^2$, Romina Peruilh Bagolini$^2$}
\date{
	$^1$Department of Statistics and Quantitative Methods, University of Milano-Bicocca, Via Bicocca degli Arcimboldi 8, 20126 Milan, Italy, \\ \href{mailto:luca.brusa@unimib.it}{luca.brusa@unimib.it}, \href{mailto:fulvia.pennoni@unimib.it}{fulvia.pennoni@unimib.it} \\
	$^2$Department of Economics, University of Perugia, Via Alessandro Pascoli 20, 06123 Perugia, Italy, \\ \href{mailto:francesco.bartolucci@unipg.it}{francesco.bartolucci@unipg.it}, \href{mailto:romina.peruilh@unipg.it}{romina.peruilh@unipg.it}
}

\begin{document}
\maketitle

\noindent
{\bfseries Abstract}\\
A penalized maximum likelihood estimation approach is proposed for discrete-time hidden Markov models  where covariates affect the observed responses and serial dependence is considered. 
The proposed penalized maximum likelihood method addresses the issue of latent state separation that typically occurs when this model is applied to binary and categorical response variables with a limited number of categories, resulting in  extremely large estimates of the support points of the latent variable assumed with a discrete, left unspecified distribution. We also propose a cross-validation approach for jointly selecting the number of hidden states and the roughness of the penalty term.  
The proposal is illustrated through a simulation study comparing parameter estimation accuracy and computational efficiency across different estimation procedures. We also demonstrate the potential of this class of models through the analysis of longitudinal data collected during spinal anesthesia to monitor the occurrence of hypotension in patients, and we compare the results with those obtained from other standard models.


\newpage

\section{Introduction}
In many longitudinal studies, the focus lies in the evolution of a certain phenomenon of interest among groups of individuals over time, which is assessed using occasion-specific response variables. Moreover, it is often crucial to assess how covariates affect the response variables and how this influence changes over time. Some relevant issues include assessing what factors contribute to the variability in the response variables and how unobserved heterogeneities affect individual responses over time. 

Numerous models are available for the sort of analyses mentioned above, including hidden (or latent) Markov (HM) models \citep{bart:pen13, mor:21, visser2022mixture}. These models are suitable for clustering individuals, as they assume a latent process that affects the distribution of the response variables. This latent variable has a discrete distribution left unspecified \citep{bart:et22}, and the process follows a  Markov chain typically of first-order with a finite number of states and assumes local independence. 
HM models may be also formulated including individual covariates either in the measurement (sub)model, affecting directly the conditional response probabilities, or in the latent (sub)model allowing the initial and transition probabilities of the latent Markov chain to depend on the covariates through a suitable parameterization \citep{Bart:Farc:Penn:2014, maruotti2017model}. 
When individual-level covariates are assumed to influence the latent state dynamics, shaping the transition process, it is typically considered that measurement errors are present \citep{bart:pen13}.
When the response variables depend on covariates which are subject specific, the latent variable accounts for the so-called unobserved heterogeneity, which refers to differences in individuals' responses that cannot be explained by the observed covariates. In other words, even after accounting for all measurable individual characteristics, there may still be variability in the data due to unmeasured or latent factors. This is a crucial aspect to consider in the analysis of longitudinal data as failing to account for it can lead to biased or incomplete conclusions. 
Moreover, in many applications, it may be of interest to consider conditional serial dependence between the responses. Including the lagged response among covariates relaxes the assumption of conditional independence, which is often deemed too restrictive in applied contexts such in capture-recapture studies \citep{bart:pen07}.

Considering categorical response variables, a convenient parameterization of the conditional distribution of the responses in a HM model is based on a generalized logit parameterization. This model specification includes parameters measuring the effect of each hidden state on the probability of success at each time occasion, as well as regression parameters for covariates and lagged response. In this way, the hidden states are easily interpretable as a tendency or propensity towards certain behaviors, and the overall model can be seen as a generalization of the dynamic logit model \citep{hsia:03} with time-varying random effects.

Maximum likelihood estimation of the models at issue is carried out in a relatively simple way using  the expectation-maximization (EM) algorithm \citep{demp77}.  
At each maximization step, the optimization involving the covariate parameters is based on the Newton-Raphson (NR) iterative algorithm, which is run starting from the results of the corresponding expectation step of the EM algorithm.
To assess the precision of the maximum likelihood parameter estimates, asymptotic standard errors can be obtained in the usual way through the observed information matrix, which is calculated using numerical methods \citep{bart:farc:09}. 
The model formulation based on discrete latent variables allows for simpler inferential procedures compared to other latent variable model specifications where the random effects have a continuous distribution, such as those proposed by \cite{cagnone2009latent}. 
However, it may happen that the parameters estimates of the support points related to the hidden states assume extreme values due to a sort of separation problem.

The separation problem is well-known and has been extensively studied in many statistical models for categorical response variables, such as multiple logistic regression; see among others, \citet{mans:etal:17} and \citet{clar:etal:23}.
In these contexts it typically arises when covariates perfectly predict the response variable, which may result in infinite estimates for some regression coefficients. The causes of separation are well-established and primarily include small sample sizes, rare outcomes, and highly correlated covariates \citep{mans:etal:17}.

In the context of HM models, instead, separation primarily affects the support points of latent states, resulting in extremely large estimates. Consequently, the estimated covariate effects may become negligible. 
To address this issue, in the present paper we propose a penalized likelihood method, based on the squared distance between each support point and their average, to prevent overly large estimates of the support points and ensure reliable statistical inference.
This approach was introduced by \cite{good:gask71} and later applied by \citet{silverman82} for estimating probability densities through the maximization of a penalized likelihood function.
We demonstrate that the resulting constrained maximization can be performed by properly implementing the M-step of the EM algorithm mentioned above. 
Additionally, we propose a procedure, inspired by that in \cite{smyth00}, for cross-validating the likelihood in model-based probabilistic clustering contexts. Specifically, we introduce cross-validation to jointly select the number of states of the hidden chain and the roughness of the penalty \citep{scott78}.

To evaluate the asymptotic properties of the proposed estimator, we conduct a simulation study. We compare the performance of penalized and standard maximum likelihood estimation in terms of the accuracy of the estimates and computational time. 
Additionally, we present the results of the cross-validation approach to select the best value of the penalization parameter.
We also apply the model to data related to the occurrence of hypotension measured every five minutes during spinal anesthesia (SA), which still remains an open field of research as shown in \cite{radwan:24}.  Our objective is to identify subgroups of patients exhibiting distinct developmental patterns of hypotension and assess the direct effects of time-fixed and time-varying covariates.
This application also offers the possibility to compare the results of the HM model with those provided by \cite{aktas14}, who used the generalized linear mixed model \citep{fitzmaurice2012applied} and the generalized estimating equations \citep{liang:86}.

The reminder of the paper is organized as follows. In Section \ref{sec:model}, we recall the model formulation and maximum likelihood estimation. In Section \ref{sec:pen}, we present the proposed penalization method and the estimation procedure. In Section \ref{sec:sim}, we show the results of the simulation study, and in Section \ref{sec:app}, we illustrate the results of the application. Finally,  in Section \ref{sec:con}, we provide some concluding remarks. 
Appendix \ref{app} includes additional details regarding the simulation results. 
The estimation procedures are implemented by extending the functions of the \texttt{LMest} package \citep{bart:et17} in the software \textsf{R} \citep{core:24}.

\section{Model formulation}\label{sec:model}
Let $\bm{Y}_i = (Y_i^{(1)}, \ldots, Y_i^{(T)})'$ and $\bm{X}_i = (\bm{X}_i^{(1)}, \ldots, \bm{X}_i^{(T)})'$,  where $Y_i^{(t)}$  denotes the univariate binary response variable and $\bm{X}_i^{(t)}$ the vector of individual covariates for subject $i$ at time $t$, for $i=1,\ldots,n$ and $t=1,\ldots,T$. 
In addition, let $\bm{U}_i = (U_i^{(1)}, \ldots, U_i^{(T)})'$ be a latent process assumed as a first-order Markov chain with a finite number of hidden states $\{1, \ldots, k\}$. 
In the following, $\bm{y} = (y^{(1)}, \ldots, y^{(T)})'$, $\bm{u} = (u^{(1)}, \ldots, u^{(T)})'$, and $\bm{x} = (\bm{x}^{(1)}, \ldots, \bm{x}^{(T)})'$ denote realizations of $\bm{Y}_i$, $\bm{X}_i$, and $\bm{U}_i$, respectively.
The HM model is characterized by two (sub)models: ($i$) the measurement (sub)model, whose probability mass (or density) function is denoted as $p_{\bl{Y}_i | \bl{U}_i, \bl{X}_i}(\bm{y} | \bm{u}, \bm x)$ and represents the distribution of the response vector $\bm{Y}_i$ given the latent process $\bm{U}_i$ and the observable covariates $\bm{X}_i$, and ($ii$) the latent (sub)model, whose probability mass function is denoted as  $p_{\bl{U}_i}(\bm{u})$ and represents the distribution of the latent process. 
In this formulation, the covariates are assumed to affect solely the measurement (sub)model. 
In the following, we also refer to the manifest distribution of the response variable given the covariates as
\begin{equation*}
	p_{\bl{Y}_i | \bl{X}_i} (\bm{y} | \bm{x}) = \sum_{\bm{u}} p_{\bl{Y}_i | \bl{U}_i, \bl{X}_i} (\bm{y} | \bm{u}, \bm{x}) p_{\bl{U}_i} (\bm{u}).
\end{equation*}

Parameters related to the latent sub-model do not depend on covariates and they are: ($i$) the a priori probability of each state, denoted as $\pi_{u} = p(U_i^{(1)} = u)$, $u = 1, \ldots, k$, satisfying $\pi_{u} \geq 0$ and $\sum_{u=1}^{k}\pi_{u} = 1$, and ($ii$) the transition probabilities among hidden states, denoted as $\pi_{u|\bar{u}} = p(U_i^{(t)} = u | U_i^{(t-1)} = \bar{u})$, $\bar{u}, u = 1, \ldots, k$, $t = 1, \ldots, T$, satisfying $\pi_{u|\bar{u}} \geq 0$ and $\sum_{u=1}^{k}\pi_{u|\bar{u}} = 1$. 
Some additional constraints are required for identifiability. 
Hidden states correspond to different subpopulations and they are associated to different levels of the effect of the unobserved covariates on the response variable. In this way, unobserved heterogeneity is taken into account dynamically since each individual may move between hidden states over time.

According to the local independence assumption, the response variables are generally assumed to be conditionally independent given $\bm{U}_i$. 
This assumption can be relaxed by including in each vector of covariates $\bm{X}_i^{(t)}$ the lagged response variable $Y_i^{(t-1)}$, allowing for serial dependence between observed responses over time.  
Suitable parameterizations can be adopted to explain the dependence of the response variable on such covariates \citep{bart:pen13}. Here we consider a parameterization in the form of a  generalized linear model \citep{mcc:nel:89, agre:02} of the following type:
\begin{equation}\label{eq:logistic}
	\log\frac{p(Y_i^{(t)} = 1 | U_i^{(t)} = u, \bm{X}_i^{(t)} = \bm{x})}{p(Y_i^{(t)} = 0 | U_i^{(t)} = u, \bm{X}_i^{(t)} = \bm{x})} = \log\frac{\phi^{(t)}_{1|u\bm{x}}}{\phi^{(t)}_{0|u\bm{x}}} = \log\frac{\phi^{(t)}_{1|u\bm{x}}}{1 - \phi^{(t)}_{1|u\bm{x}}} = \alpha_u + \bm{x}'\bm{\beta},
\end{equation}
where $\phi_{1|u\bm{x}}^{(t)}$ and $\phi_{0|u\bm{x}}^{(t)}$ denote the conditional response probabilities given the hidden state and the covariate configuration, $u = 1, \ldots, k$, $t = 1, \ldots, T$. 
Here $\bm{\beta}$ is the vector of regression parameters for the covariates and $\bm{\alpha} = (\alpha_1, \ldots, \alpha_k)$ is the vector of specific support points corresponding to each hidden state; $\alpha_u$ measures the tendency toward the success event for subjects in state $u$. To ensure model identifiability, we fix the constraints that $\alpha_1 < \alpha_2 < \ldots < \alpha_k$ and that $\alpha_1$  = 0.

One key feature of the resulting model is the estimation of the time-varying direct effect of each covariate on the response variable.
The model represents a discrete version of a random-effect logit model \citep{mccull08}, extending the dynamic logit model proposed in \cite{hsia:03}.

\subsection{Maximum likelihood estimation}\label{subsec:mle}
Maximum likelihood estimation of the model parameters, collected in the vector $\bm{\theta}$, is based on a sample of $n$ observations and performed through the EM algorithm \citep{demp77}. 
This algorithm maximizes the observed-data log-likelihood function: 
\[
	\ell(\bm{\theta}) = \sum_{i=1}^{n} \log p_{\bl{Y}_i | \bl{X}_i} (\bm{y} | \bm{x}).
\]
In order to compute the manifest distribution of the observable responses, the Baum and Welch recursion is used \citep{baum:petr:66}. 
The complete-data log-likelihood function can be expressed as 
\[
	\ell^*(\bm{\theta}) = \sum_{i=1}^{n} \log p_{\bl{Y}_i, \bl{U}_i | \bl{X}_i} (\bm{y}, \bm{u} | \bm{x}).
\]

It is often convenient to disentangle this function into three components referred to the model parameters:
\begin{equation}\label{eq:comp.loglik}
	\begin{aligned}
		\ell^*(\bm{\theta}) 
			&= \sum_{i=1}^{n}\sum_{u=1}^{k}\sum_{t=1}^{T} z_{iu}^{(t)} \log p_{{\scriptscriptstyle Y}_i^{(t)} | {\scriptscriptstyle U}_i^{(t)}, \bl{X}_i^{(t)}} (y^{(t)} | u^{(t)}, \bm{x}^{(t)}) \\
			&+ \sum_{i=1}^{n}\sum_{u=1}^{k} z_{iu}^{(1)} \log\pi_{u} + \sum_{i=1}^{n}\sum_{u=1}^{k}\sum_{\bar{u}=1}^{k} z_{i\bar{u}u} \log\pi_{u|\bar{u}}, 
	\end{aligned}
\end{equation}
where $z_{iu}^{(t)}$ is an indicator function equal to 1 if subject $i$ is in hidden state $u$ at time $t$, and $z_{i\bar{u}u} = \sum_{t>1} z_{i\bar{u}}^{(t-1)} z_{iu}^{(t)}$ denotes the number of times subject $i$ moves from state $\bar{u}$ to state $u$. 

After a proper initialization of the model parameters, the EM algorithm alternates between the following two steps until convergence: 
\begin{itemize}
	\item[($i$)] \textbf{expectation step}: this step computes the conditional expected value of $\ell^*(\bm{\theta})$ given the observed data and the estimates of the parameters from the previous step. It is obtained by substituting the variables $z_{iu}^{(t)}$ and $z_{i\bar{u}u}$ in \eqref{eq:comp.loglik} with their expected values:
	\begin{gather}
		\hat{z}_{iu}^{(t)} = q_{{\scriptscriptstyle U}_i^{(t)} | \bl{X}_i,\bl{Y}_i} (u | \bm{x},\bm{y}), \label{eq:posterior}\\
		\hat{z}_{i\bar{u}u} = \sum_{t>1} q_{{\scriptscriptstyle U}_i^{(t-1)}, {\scriptscriptstyle U}_i^{(t)} | \bl{X}_i, \bl{Y}_i} (\bar{u}, u |  \bm{x},\bm{y}).
	\end{gather}
	Here $q_{{\scriptscriptstyle U}_i^{(t)} | \bl{X}_i,\bl{Y}_i} (u | \bm{x},\bm{y})$ denotes the conditional probability of the latent variable at time $t$ given the response and covariate configurations, while $q_{{\scriptscriptstyle U}_i^{(t-1)}, {\scriptscriptstyle U}_i^{(t)} | \bl{X}_i, \bl{Y}_i} (\bar{u}, u |  \bm{x},\bm{y})$ represents the conditional distribution of the latent variables at times $t$ and $t-1$ given the response and covariate configurations. 
	The computation of these posterior distributions is computationally challenging and requires suitable backward recursions; see \citet[Chapter 5]{bart:pen13} for more details.
	\item[($ii$)] \textbf{maximization step}: this step  maximizes the expectation of $\ell^*(\bm{\theta})$ and updates the value of the parameters.  Explicit solutions are available for the initial and transition probabilities:
	\begin{gather*}
		\hat{\pi}_u = \frac{\sum_{i=1}^{n} \hat{z}_{iu}^{(t)}}{n},\\
		\hat{\pi}_{u|\bar{u}} = \frac{\sum_{i=1}^{n} \hat{z}_{i\bar{u}u}}{\sum_{i=1}^{n}\sum_{u=1}^{k} \hat{z}_{i\bar{u}u}}.
	\end{gather*}
	Solutions for the remaining parameters $\bm{\alpha}$, and $\bm{\beta}$, collected in the vector $\bm{\theta}_1 = (\bm{\alpha}', \bm{\beta}')'$, are obtained using a Newton-Raphson algorithm through the following iterative step:
	\begin{equation}\label{eq:NewtonRaphson}
		\bm{\theta}_1^{(h+1)} = \bm{\theta}_1^{(h)} + \bm{F}^{-1}\bm{s},
	\end{equation}
	with $\bm{s} = \frac{\partial \ell^*(\bm{\theta})}{\partial \bm{\theta}_1}$ and $\bm{F} = \frac{\partial^2 \ell^*(\bm{\theta})}{\partial \bm{\theta}_1^2}$, where  $h$ refers to the $h$-iteration of the Newton-Raphson algorithm.
\end{itemize}

To address the problem of multi-modality of the log-likelihood function, parameter initialization typically relies on a multi-start strategy, using deterministic and random rules \citep{bart:pen13}. The global maximum is assumed to correspond to the highest value of the log-likelihood function. 
Regarding convergence of the algorithm, two common rules are employed: the algorithm is stopped when either the relative change in the log-likelihood function between iterations falls below a predefined threshold $\varepsilon_1$ or the change in the model parameters between iterations falls below a predefined threshold $\varepsilon_2$:
\begin{equation*}
	\frac{\ell(\bm{\theta}^{(h)}) - \ell(\bm{\theta}^{(h-1)})}{\lvert \ell(\bm{\theta}^{(h)}) \rvert} < \varepsilon_1 \quad\text{and}\quad \max_{s} \lvert \theta_s^{(h)} - \theta_s^{(h-1)} \rvert < \varepsilon_2.
\end{equation*}

Standard errors for the parameter estimates can be obtained using the numerical method proposed in \cite{bart:farc:09}, which involves taking the negative of the first derivative of the score vector at convergence. The score vector is, in turn, obtained as the first derivative of the expected value of the complete data log-likelihood, as suggested in \cite{oake:99}; see also \cite{penn:14}.
Concerning model identifiability, we note, as pointed out in \cite{bart:pen13}, that the model is globally identifiable only up to a permutation of the latent states. However, local identifiability can  be assessed by examining the observed information matrix and verifying whether it is of full rank \citep{mchu:56, roth:71, good:74b}.

Once the model parameters are estimated, prediction of the hidden state for each unit at each time occasion based on the observed data is performed through local decoding. This estimation requires to find the most likely hidden state for every time occasion. 
Considering the estimated posterior probabilities $\hat{z}_{iu}^{(t)}$, the predicted state is found by maximizing these probabilities.
The entire sequence of predicted hidden states produced by the local decoding is $\hat{\bm u}(\bm y) = (\hat{u}^{(1)}(\bm{y}),\ldots,\hat{u}^{(T)}(\bm{y}))'$.

Considering the posterior probabilities $\hat{z}_{iu}^{(t)}$ in expression (\ref{eq:posterior}), obtained as the results of the E-step of the estimation algorithm, a weighted mean of the values of $\hat\alpha_{u}$ for each subject $i$ can be obtained as
\begin{equation}
	\hat{\bar{\alpha}}_i^{(t)} = \sum_{u=1}^{k} \hat\alpha_{u} \:\hat{z}_{iu}^{(t)}, 
\end{equation}
where $\hat{\bar{\alpha}}_i^{(t)}$ can be referred to as  the individual latent effects mean.

\section{Penalized maximum likelihood estimation}\label{sec:pen}
We consider the generalized logit parameterization of the measurement (sub)model as expressed in Equation \eqref{eq:logistic}. In the case of widely separated hidden states, significant differences in the values of $\alpha_u$ are observed. 
This may result in ($i$) higher relevance of one or more states compared to others, and ($ii$) reduced importance of the available covariates, whose estimated effects may become negligible or insignificant.
In the following sections, we illustrate the challenges of separation in HM models through a practical example (Section \ref{subsec:example}), present the penalization method and the related estimation procedure (Section \ref{subsec:penalization}), and describe a cross-validation approach to select the roughness of the penalization (Section \ref{subsec:cross}).

\subsection{A motivating example}\label{subsec:example}
Considering the model introduced in Section \ref{sec:model}, we generate data with a binary response variable and four covariates, one of which is the lagged response. The regression coefficients for the covariates are $\bm{\beta} = [1, -1, 1, 1]'$. 
We assume $k=3$ latent states, with support points $\bm{\alpha} = [-20, -5, 5]'$. The initial probabilities are set to $1/3$ for each latent state, and the transition probability matrix has diagonal values of 0.750 and off-diagonal values of 0.125. 
The dataset consists of 250 subjects over 10 time points. The HM model is fitted using the standard maximum likelihood approach described in Section \ref{subsec:mle}, and the estimated model parameters are reported in Table \ref{tab:example}. 

The estimated values of $\alpha_u$ are highly distant from each other, resulting in widely separated latent states. This determines a significant imbalance in the importance of different latent states, as confirmed by the allocation of subjects using local decoding. For example, at the initial time point, 159 subjects are allocated to the 1st latent state and 91 to the 3rd, with none assigned to the 2nd. This contrasts with the true data generation process, where 88 subjects were allocated to the 1st state, 71 to the 2nd, and 91 to the 3rd. Additionally, based on the estimated parameters, subjects in the 1st  state consistently have a conditional response probability $\phi^{(t)}_{1|u\bm{x}}$ equal to 0, while those in the 3rd have a probability of 1. This indicates that the covariates have no effect on the estimated conditional probabilities and on the posterior state allocation. 

It is also worth noting that the estimated values of $\alpha_u$ differ substantially from the true values used in data generation. Notably, while the true support points included two negative values and one positive, the estimated values show the opposite pattern, with two positive and only one negative. This indicates poor accuracy in parameter estimation. Similar issues arise with the estimated regression coefficients, $\beta_j$, which also differ significantly from the values used to generate the data. Furthermore, none of the covariates are found to be statistically significant, with all $p$-values exceeding 0.25.
\begin{table}[t!]
	\centering
	\caption{Maximum likelihood estimates of the parameters of the HM model with three latent states for data simulated with $\bm{\beta} = [1, -1, 1, 1]'$ and $\bm{\alpha} = [-20, -5, 5]'$}
	\vspace{-0.1cm}
	\begin{tabular}{lrrrrrrr}
		\toprule
						& \multicolumn{1}{c}{$\hat\alpha_1$} & \multicolumn{1}{c}{$\hat\alpha_2$} & \multicolumn{1}{c}{$\hat\alpha_3$} & \multicolumn{1}{c}{$\hat\beta_1$} & \multicolumn{1}{c}{$\hat\beta_2$} & \multicolumn{1}{c}{$\hat\beta_3$} & \multicolumn{1}{c}{$\hat\beta_4$} \\
		\midrule
		Estimate		& -250.126	& 146.804	& 457.960	& 24.142	& -80.211	& 56.686	& 103.479	\\
		Standard error 	& -			& -			& -			& 23.283	& 69.426	& 50.905	& 90.144	\\
		$p$-value 		& -			& -			& -			& 0.300		& 0.248		& 0.265		& 0.251		\\
		\bottomrule
	\end{tabular}
	\label{tab:example}
\end{table}

\subsection{Penalization term}\label{subsec:penalization}
To solve the problem of latent state separation, we propose a penalization term $\mathcal{A}$ aimed at reducing the separation between hidden states. 
With the purpose of decreasing the relative distance among the elements of vector $\bm{\alpha}$, we define $\mathcal{A}$ as
\begin{equation*}
	\mathcal{A} = \sum_{u=1}^{k} (\alpha_u - \bar{\alpha})^2 = \sum_{u=1}^{k} \alpha_u^2 - k \bar{\alpha}^2,
\end{equation*}
where $\bar{\alpha} = \frac{1}{k}\sum_{u=1}^{k}\alpha_u$.  
In matrix notation
\begin{equation*}
	\mathcal{A} = \bm{\alpha}' \bm{J} \bm{\alpha}, 
\end{equation*}
where $\bm{J} = \bm{I} - \frac{1}{k} \bm{1} \bm{1}'$ is a $k \times k$ matrix, $\bm{I}$ is the identity matrix of size $k$, and $\bm{1} = (1, \ldots, 1)'$ is the column vector of ones of size $k$.

The proposed penalization term is applied to both the observed data and the complete data log-likelihood functions:
\begin{gather*}
	\tilde{\ell}(\bm{\theta}) = \ell(\bm{\theta}) - \lambda \mathcal{A}, \\
	\tilde{\ell}^*(\bm{\theta}) = \ell^*(\bm{\theta}) - \lambda \mathcal{A},
\end{gather*}
where $\lambda \in \mathbb{R}^{+}$ is a tuning parameter that determines the strength of the penalty term.
In this way, by maximizing the penalized log-likelihood function, we seek for a balance between the model that best fits the data and the model with minimal distance between the support points of the hidden states. 

Maximum likelihood estimation of the penalized model parameters is performed again using the EM algorithm; the expectation step remains unaltered, while the maximization step requires revising the Newton-Raphson iterative step expressed in Equation \eqref{eq:NewtonRaphson}. In particular, the formulas to compute the first and second derivatives of the complete-data log-likelihood function with respect to the vector $\bm{\alpha}$ are updated as follows:
\begin{equation*}
	\tilde{\bm{s}}_{\bm{\alpha}} = \frac{\partial \tilde{\ell}^*(\bm{\theta})}{\partial \bm{\alpha}} = \frac{\partial \ell^*(\bm{\theta})}{\partial \bm{\alpha}} - 2\lambda \bm{J} \bm{\alpha} = \bm{s}_{\bm{\alpha}} - 2\lambda \bm{J} \bm{\alpha},
\end{equation*}
and
\begin{equation*}
	\tilde{\bm{F}}_{\bm{\alpha}\bm{\alpha}} = \frac{\partial^2 \tilde{\ell}^*(\bm{\theta})}{\partial\bm{\alpha}^2} = \frac{\partial^2 \ell^*(\bm{\theta})}{\partial\bm{\alpha}^2} - 2\lambda \bm{J} = \bm{F}_{\bm{\alpha}\bm{\alpha}} - 2\lambda \bm{J},
\end{equation*}
where $\bm{s}_{\bm{\alpha}}$ denotes the sub-vector of $\bm{s}$ and $\bm{F}_{\bm{\alpha}\bm{\alpha}}$ the sub-matrix of $\bm{F}$ corresponding to the derivative (first and second, respectively) with respect to $\bm{\alpha}$. The remaining components of $\tilde{\bm{s}}$ and $\tilde{\bm{F}}$ remain instead unchanged.

\subsection{Cross-validation}\label{subsec:cross}
The tuning parameter $\lambda$ regulates the intensity of penalization in the model. When $\lambda = 0$ no penalty is applied, resulting in estimates identical to those of the maximum likelihood method. 
Conversely, as $\lambda$ approaches infinity, the values of the support points $\alpha_u$ tend toward 0, and the relative importance of the unobserved heterogeneity tends toward zero.
A reliable method to select the value of $\lambda$ is cross-validation, which is commonly employed in other contexts, see, among others,  \cite{kohavi95, smyth00, hastie09, bates23}.

In this work, we employ a cross-validation approach to jointly select both the value of the penalization parameter $\lambda$ and the number of hidden states $k$. Denoting the data matrix as $D$, and following the procedure proposed by \citet{smyth00} and \citet{bart:etal:17} for the selection of the number of components in the context of finite mixture models \citep{mcla:peel:00} and HM models, respectively, we consider $M$ partitions of the data $(D \backslash S_m, S_m)_{m = 1, \ldots, M}$. 
For the $m$-th partition, the model is estimated on the data subset $D \backslash S_m$, providing parameters estimates $\bm{\theta}^{(k, \lambda)}(D \backslash S_m)$. Let  $\ell\left( \bm{\theta}^{(k, \lambda)}(D \backslash S_m) \ | \ S_m \right)$ denote the (possibly penalized) log-likelihood function where the model parameters are estimated on the training data $D \backslash S_m$ but, the log-likelihood is evaluated on the test data $S_m$. Finally, the cross-validated likelihood is defined as
\begin{equation*}
	\ell_{\textsc{cv}} = \frac{1}{M} \sum_{m = 1}^{M} \ell\left( \bm{\theta}^{(k, \lambda)}(D \backslash S_m) \ | \ S_m \right).
\end{equation*}

\section{Simulation study}\label{sec:sim}
A simulation study is designed to evaluate the quality of the proposed penalized maximum likelihood approach. 
In the following, we illustrate the simulation design (Section \ref{subsec:sim0}) and summarize the main results (Sections \ref{subsec:sim1} and \ref{subsec:sim2}).

\subsection{Simulation design}\label{subsec:sim0}
Forty different scenarios are considered, fixing the number of hidden states to $k = 3$, and 
varying the sample size ($n = 250, 500$), the number of time occasions ($T = 10, 20$), the state persistence (high or low), and the state separation for which we suppose five different values of the support points $\alpha_u$, denoted as $\bm{\alpha}^j$, $j = 1, \ldots, 5$ defined as: $\bm{\alpha}^{1} = [-3, 0, 3]'$, $\bm{\alpha}^{2} = [-6, 0, 6]'$, $\bm{\alpha}^{3} = [-6, -3, 0]'$, $\bm{\alpha}^{4} = [-20, -5, 5]'$, and $\bm{\alpha}^{5} = [-40, -15, 0]'$. 
In particular, two different transition patterns are considered by  setting the following  matrices of transition probabilities characterized by higher or lower levels of persistence, respectively:
\begin{equation*}
	\bm{\Pi}^{\text{high}} = \begin{bmatrix}
		0.900 & 0.050 & 0.050 \\
		0.050 & 0.900 & 0.050 \\
		0.050 & 0.050 & 0.900
	\end{bmatrix} \qquad \text{and} \qquad \bm{\Pi}^{\text{low}} = \begin{bmatrix}
		0.750 & 0.125 & 0.125 \\
		0.125 & 0.750 & 0.125 \\
		0.125 & 0.125 & 0.750
	\end{bmatrix}.
\end{equation*}

In each scenario, four independent standard normally distributed covariates are considered for the model formulation defined in Equation  (\ref{eq:logistic}), and  the corresponding vector of regression coefficients $\bm{\beta}$ set to $(1, -1, 1, 1)'$. 

In Section \ref{subsec:sim1} the following criteria are used to evaluate the results: ($i$) mean squared error (MSE) between true and estimated model parameters, ($ii$) standard errors of the covariate regression parameters $\bm{\beta}$, and ($iii$) computational time.
In Section \ref{subsec:sim2} selection of the penalization parameter is assessed through the cross-validation approach introduced in Subsection \ref{subsec:cross}.

\subsection{Simulation results}\label{subsec:sim1}
For each of the 40 scenarios, we randomly draw 50 samples and we estimate the HM model described in Section \ref{sec:model}, using both standard and penalized approaches. We consider two values for the penalization parameter, $\lambda = 0.01$, and $\lambda = 0.05$.
To mitigate the risk of convergence to local maxima, for every sample we repeat the estimation of each model 25 times, using both deterministic and random initialization methods. Inference is then based on the solution corresponding with the highest likelihood value at convergence.

We note that, when checking model identifiability using the method introduced in Section \ref{subsec:mle}, the observed information matrix was not of full rank in fewer than 2\% of cases—that is, in 16 out of the 40 scenarios considered—and only for the estimates obtained without penalization. Therefore, it is worth mentioning that the proposed penalized maximum likelihood method is also useful for addressing identifiability issues that may arise in the context of latent variable models. The following results exclude these cases to ensure a fair comparison between standard and penalized estimation approaches.

To assess the performance of the proposal, we first consider, for every scenario, the MSE between each true model parameter $\theta$ and the corresponding estimate $\hat{\theta_s}$ obtained from  each sample $s$, $s=1, \ldots, 50$: 
\begin{equation*}
	\text{MSE}(\theta) = \frac{1}{50} \sum_{s=1}^{50} ( \theta - \hat{\theta_s} )^2. 
\end{equation*}

\begin{figure}[th!]  
	\centering
	\caption{Percentage variation in the mean squared error between true and estimated parameter for the covariate regression coefficient $\beta_4$ for the penalized estimation ($\lambda = 0.01$, in {\color{col1}pink}, and $\lambda = 0.05$, in {\color{col2}violet}) compared to the standard estimation ($\lambda = 0.00$). Each scenario is identified by sample size ($n = 250, 500$), number of time occasions ($T = 10, 20$), state persistence (H: high or L: low), and state separation ($\bm{\alpha}^1, \ldots, \bm{\alpha}^5$)}
	\includegraphics[width=.90\textwidth]{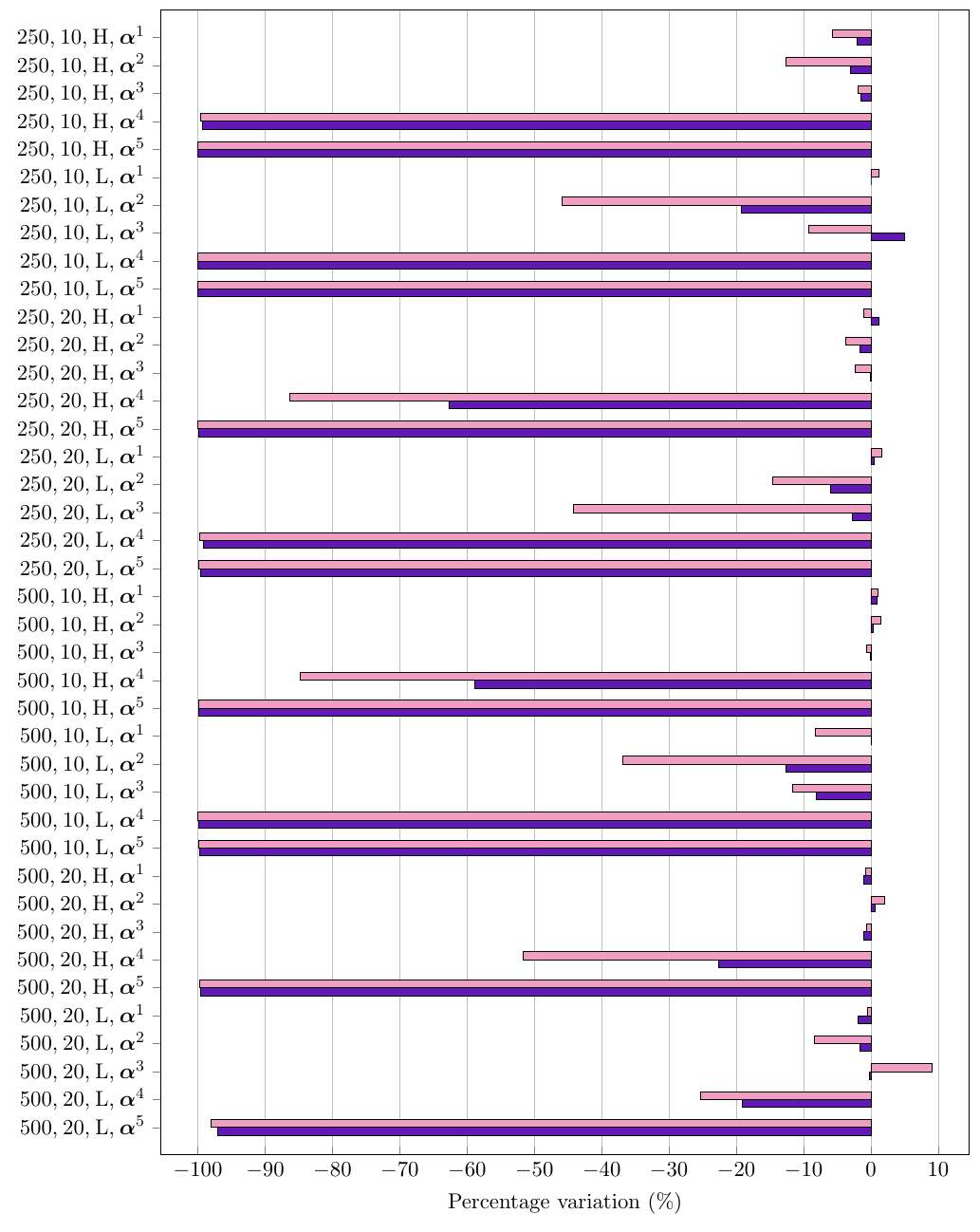}        
	\label{fig:mse_beta} 
\end{figure}
Figure \ref{fig:mse_beta} reports the percentage variation in this quantity for the covariate regression coefficient $\beta_4$, for the penalized maximum likelihood estimation ($\lambda = 0.01$ and $\lambda = 0.05$) with respect to the standard approach ($\lambda = 0.00$). 
We observe that:
\begin{itemize}
	\item in the majority of cases (31 out of 40), both values of $\lambda$ ensure a lower MSE when using the penalized estimation method. This indicates that the estimated parameter is, on average, closer to the true values, thus more accurate;
	\item scenario related to the fifth latent state separation behavior ($\bm{\alpha}^5$, characterized by highly separated states) shows the greatest improvement; the percentage decreases using penalized estimation are often exceeding $90\%$; 
	\item on the contrary, the penalization approach appears less effective under the first separation behavior ($\bm{\alpha}^1$, characterized by closely spaced states) and in cases with a high value of $T$ (20 time occasions); 
	\item only in four cases (out of 40) the penalized estimation method exhibits no improvement with either value of $\lambda$, showing slight increases in the MSE value.
\end{itemize}
Tables \ref{tab:beta1}, \ref{tab:beta2}, \ref{tab:beta3}, and \ref{tab:beta4} in Appendix \ref{app} provide additional results related to the MSE for each regression coefficient $\beta_1$, $\beta_2$, $\beta_3$, and $\beta_4$.

\begin{table}[t!]
	\centering
	\caption{Percentage variation in the standard errors of the covariates regression parameters for the penalized estimation ($\lambda = 0.01$ and $\lambda = 0.05$) with respect to the standard maximum likelihood estimation ($\lambda = 0.00$). Each scenario is identified by sample size ($n = 250, 500$), number of time occasions ($T = 10, 20$), state persistence (High or Low), and state separation ($\bm{\alpha}^1, \ldots, \bm{\alpha}^5$)}
	\small
	\begin{tabular}{llllrrrrr}
		\toprule
		& & & & \multicolumn{1}{c}{$\bm{\alpha}^1$}	& \multicolumn{1}{c}{$\bm{\alpha}^2$}	& \multicolumn{1}{c}{$\bm{\alpha}^3$}	& \multicolumn{1}{c}{$\bm{\alpha}^4$}	& \multicolumn{1}{c}{$\bm{\alpha}^5$} \\
		\cmidrule(lr){5-9}
		$n$	& $T$	& $\bm{\Pi}$	& $\lambda$	& \multicolumn{5}{c}{Percentage variation (\%)}	\\
		\midrule
		\multirow{8}[5]{*}{$250$}	& \multirow{4}[2]{*}{$10$}	& \multirow{2}{*}{High}	& $0.01$	& $+$1.76	& $-$42.86	& $-$71.36	& $-$60.62	& $-$96.95	\\
									& 							& 						& $0.05$	& $-$4.64	& $-$51.13	& $-$74.65	& $-$72.18	& $-$98.03	\\
		\cmidrule(lr){3-9}
									& 							& \multirow{2}{*}{Low}	& $0.01$	& $+$0.47	& $-$46.90	& $-$52.64	& $-$94.66	& $-$95.15	\\
									& 							& 						& $0.05$	& $-$5.61	& $-$58.80	& $-$60.19	& $-$96.48	& $-$97.15	\\
		\cmidrule(lr){2-9}
									& \multirow{4}[2]{*}{$20$}	& \multirow{2}{*}{High}	& $0.01$	& $-$0.02	& $-$16.37	& $-$67.37	& $-$37.15	& $-$94.13	\\
									& 							& 						& $0.05$	& $-$0.53	& $-$25.53	& $-$73.59	& $-$54.55	& $-$95.29	\\
		\cmidrule(lr){3-9}
									& 							& \multirow{2}{*}{Low}	& $0.01$	& $+$0.08	& $-$41.99	& $-$84.63	& $-$75.16	& $-$88.37	\\
									& 							& 						& $0.05$	& $-$4.05	& $-$54.27	& $-$85.65	& $-$83.28	& $-$93.23	\\
		\midrule
		\multirow{8}[5]{*}{$500$}	& \multirow{4}[2]{*}{$10$}	& \multirow{2}{*}{High}	& $0.01$	& $+$6.51	& $-$23.85	& $-$69.72	& $-$73.49	& $-$95.53	\\
									& 							& 						& $0.05$	& $+$1.73	& $-$33.50	& $-$73.03	& $-$80.37	& $-$97.25	\\
		\cmidrule(lr){3-9}
									& 							& \multirow{2}{*}{Low}	& $0.01$	& $+$7.14	& $-$62.82	& $-$66.49	& $-$83.14	& $-$93.51	\\
									& 							& 						& $0.05$	& $+$0.25	& $-$70.63	& $-$68.37	& $-$88.58	& $-$95.76	\\
		\cmidrule(lr){2-9}
									& \multirow{4}[2]{*}{$20$}	& \multirow{2}{*}{High}	& $0.01$	& $+$1.58	& $-$2.80	& $-$31.01	& $-$31.79	& $-$92.53	\\
									& 							& 						& $0.05$	& $+$0.53	& $-$11.47	& $-$41.63	& $-$41.52	& $-$95.30	\\
		\cmidrule(lr){3-9}
									& 							& \multirow{2}{*}{Low}	& $0.01$	& $+$27.81	& $+$0.92	& $-$0.19	& $-$41.66	& $-$84.31	\\
									& 							& 						& $0.05$	& $+$1.53	& $-$16.50	& $-$19.00	& $-$57.98	& $-$91.28	\\
		\bottomrule
	\end{tabular}
	\label{tab:sim2}
\end{table}

A similar behavior is observed considering the estimated standard errors of the regression parameters, computed as explained in Section \ref{subsec:mle}. Table \ref{tab:sim2} reports the percentage variation in these values for the penalized estimation compared to the standard maximum likelihood estimation. 
In most cases, penalization reduces the standard errors; specifically, this reduction occurs in 32 out of 40 cases with $\lambda = 0.01$ and in 36 out of 40 cases with $\lambda = 0.05$. 
As noted earlier, the proposed approach is less effective under the first state separation behavior ($\bm{\alpha}^1$). In all other cases, the percentage decrease is considerable, often reaching very high values. 
\begin{table}[t!]
	\centering
	\caption{Percentage variation in the computational time for the penalized estimation ($\lambda = 0.01$ and $\lambda = 0.05$) with respect to the standard estimation ($\lambda = 0.00$). Each scenario is identified by sample size ($n = 250, 500$), number of time occasions ($T = 10, 20$), state persistence (High or Low), and state separation ($\bm{\alpha}^1, \ldots, \bm{\alpha}^5$)}
	\small
	\begin{tabular}{llllrrrrr}
		\toprule
		& & & & \multicolumn{1}{c}{$\bm{\alpha}^1$}	& \multicolumn{1}{c}{$\bm{\alpha}^2$}	& \multicolumn{1}{c}{$\bm{\alpha}^3$}	& \multicolumn{1}{c}{$\bm{\alpha}^4$}	& \multicolumn{1}{c}{$\bm{\alpha}^5$} \\
		\cmidrule(lr){5-9}
		$n$	& $T$	& $\bm{\Pi}$	& $\lambda$	& \multicolumn{5}{c}{Percentage variation (\%)}	\\
		\midrule
		\multirow{8}[5]{*}{$250$}	& \multirow{4}[2]{*}{$10$}	& \multirow{2}{*}{High}	& $0.01$	& $+$5.27	& $-$8.52	& $-$3.52	& $-$2.64	& $-$48.10 	\\
									& 							& 						& $0.05$	& $-$0.80 	& $-$7.41	& $+$7.86	& $-$2.28 	& $-$48.64 	\\
		\cmidrule(lr){3-9}
									& 							& \multirow{2}{*}{Low}	& $0.01$	& $+$5.89 	& $-$4.60 	& $-$4.57 	& $-$20.97 	& $-$34.16 	\\
									& 							& 						& $0.05$	& $-$1.11 	& $-$4.73	& $-$17.91 	& $-$28.55 	& $-$52.52 	\\
		\cmidrule(lr){2-9}
									& \multirow{4}[2]{*}{$20$}	& \multirow{2}{*}{High}	& $0.01$	& $+$2.20 	& $+$7.19 	& $-$4.56 	& $+$7.29 	& $-$21.32 	\\
									& 							& 						& $0.05$	& $+$6.33 	& $-$2.68	& $-$10.61 	& $+$2.40 	& $-$46.57 	\\
		\cmidrule(lr){3-9}
									& 							& \multirow{2}{*}{Low}	& $0.01$	& $-$4.89 	& $-$5.65	& $-$1.77 	& $-$23.89 	& $-$20.07 	\\
									& 							& 						& $0.05$	& $-$2.73 	& $-$14.61	& $-$7.05 	& $-$44.09 	& $-$29.43 	\\
		\midrule
		\multirow{8}[5]{*}{$500$}	& \multirow{4}[2]{*}{$10$}	& \multirow{2}{*}{High}	& $0.01$	& $+$1.39 	& $+$2.11	& $-$5.74 	& $+$10.11	& $-$42.53 	\\
									& 							& 						& $0.05$	& $-$1.29 	& $-$4.85 	& $-$9.40 	& $-$5.09	& $-$53.57 	\\
		\cmidrule(lr){3-9}
									& 							& \multirow{2}{*}{Low}	& $0.01$	& $+$8.14 	& $-$3.88  	& $+$18.30 	& $-$30.40	& $-$43.02 	\\
									& 							& 						& $0.05$	& $+$5.35 	& $-$14.62 	& $+$10.65 	& $-$49.29	& $-$60.45 	\\
		\cmidrule(lr){2-9}
									& \multirow{4}[2]{*}{$20$}	& \multirow{2}{*}{High}	& $0.01$	& $-$0.21 	& $-$2.41  	& $+$0.31 	& $-$8.63	& $-$41.77 	\\
									& 							& 						& $0.05$	& $-$3.47 	& $-$10.17 	& $-$5.98 	& $+$25.54	& $-$64.56 	\\
		\cmidrule(lr){3-9}
									& 							& \multirow{2}{*}{Low}	& $0.01$	& $-$16.22 	& $-$4.55  	& $-$7.49 	& $+$0.08	& $-$18.55 	\\
									& 							& 						& $0.05$	& $-$16.98 	& $-$12.04 	& $+$16.15 	& $-$23.35	& $-$39.58 	\\
		\bottomrule
	\end{tabular}
	\label{tab:sim3}
\end{table}

Finally, Table \ref{tab:sim3} presents the percentage variation in computational time for the penalized estimation with respect to the standard approach. Simulations are performed using a \textit{Standard\_D8ds\_v5} virtual machine with 8 vCPUs and 32 GB of RAM. 
Estimation with the penalty approach almost always reduces the average computational time, particularly in scenarios related to the fifth state separation behavior ($\bm{\alpha}^5$).

\subsection{Performance of cross-validation}\label{subsec:sim2}
The penalized estimation approach consistently shows strong performance across a wide range of scenarios. However, it is not possible to identify a single $\lambda$ value that performs best across all scenarios, as the optimal value varies from one setting to another. Therefore, in this section, we present the results obtained by selecting $\lambda$ through the cross-validation method discussed in Section \ref{subsec:cross}. In particular, for each sample, the data is split into $M=10$ equal parts (10-fold cross-validation). For each value of $\lambda \in \{ 0.00, 0.01, 0.05 \}$, the HM model is trained on 9 of these folds and tested on the remaining one; this process is repeated 10 times, each time using a different fold as the test set. Finally the cross-validated log-likelihood is computed as explained in Section \ref{subsec:cross}. The optimal value of $\lambda$ is the one corresponding to the highest value of the cross-validated log-likelihood.

We compare the results obtained through cross-validation with those derived from the standard estimation procedure. 
In particular, Table \ref{tab:simcv} reports the percentage change in MSE (as defined in Section \ref{subsec:cross}) for parameter $\beta_4$ when using the cross-validated estimation method compared to the value achieved with the classical estimation approach. 
\begin{table}[t!]
	\centering
	\caption{Percentage variation in the mean squared error between true and estimated model parameter for the covariate regression coefficient $\beta_4$ for the estimation selected using the cross-validation approach compared to the standard estimation ($\lambda = 0.00$)}
	\small
	\begin{tabular}{lllrrrrr}
		\toprule
		& & & \multicolumn{1}{c}{$\bm{\alpha}^1$}	& \multicolumn{1}{c}{$\bm{\alpha}^2$}	& \multicolumn{1}{c}{$\bm{\alpha}^3$}	& \multicolumn{1}{c}{$\bm{\alpha}^4$}	& \multicolumn{1}{c}{$\bm{\alpha}^5$} \\
		\cmidrule(lr){4-8}
		$n$	& $T$	& $\bm{\Pi}$	& \multicolumn{5}{c}{Percentage variation (\%)}	\\
		\midrule
		\multirow{4}[5]{*}{$250$}	& \multirow{2}[2]{*}{$10$}	& high	& $-$0.40	& 0.00		& $-$1.06	& $-$60.95	& $-$91.96	\\
		\cmidrule(lr){3-8}
									& 							& low	& $-$0.48	& $-$0.13	& $+$0.10	& $-$92.58	& $-$92.63	\\
		\cmidrule(lr){2-8}
									& \multirow{2}[2]{*}{$20$}	& high	& $+$0.07	& $-$0.05	& $-$0.06	& $-$12.38	& $-$91.59	\\
		\cmidrule(lr){3-8}
									& 							& low	& $+$0.41	& $+$0.03	& $-$10.35	& $-$73.81	& $-$78.18	\\
		\midrule
		\multirow{4}[5]{*}{$500$}	& \multirow{2}[2]{*}{$10$}	& high	& $-$5.34	& $-$8.87	& $+$0.38	& $-$0.01	& $-$75.38	\\
		\cmidrule(lr){3-8}
									& 							& low	& $-$5.70	& $-$22.72	& $-$1.42	& $-$94.70	& $-$59.43	\\
		\cmidrule(lr){2-8}
									& \multirow{2}[2]{*}{$20$}	& high	& $-$5.45	& $+$0.89	& $+$0.94	& 0.00		& $-$47.05	\\
		\cmidrule(lr){3-8}
									& 							& low	& $-$0.57	& $-$8.32	& $-$5.74	& 0.00		& $-$82.90	\\
		\bottomrule
	\end{tabular}
	\label{tab:simcv}
\end{table}

We observe that, in most cases, the cross-validation approach results in a lower MSE compared to the standard estimation method, thereby providing greater accuracy in estimating the model parameters. 
This effect is particularly pronounced in the eight scenarios associated with $\bm{\alpha}^5$, which show substantial reductions in MSE, usually exceeding 50\%. 
On the other hand, the use of cross-validation for selecting $\lambda$ is less effective in scenarios involving $\bm{\alpha}^3$: in 4 out of 8 cases, a very slight increase in MSE is observed compared to the standard estimation approach. 
Similar results are observed for the computation of the MSE for the parameters $\beta_1$, $\beta_2$, and $\beta_3$, and are available from the authors upon request.

\section{Application}\label{sec:app}
In medical contexts, unobserved heterogeneity is often present, and it is essential to account for it along with relevant covariates that can explain the observed heterogeneity among patients \citep{schlatt09}.
To address this, random effects and finite mixture models are often employed. When longitudinal data are available, the HM model can be a more informative method compared  to generalized linear models \citep{liang:86} as it can disentangle time-dependent variability and identify unobserved groups of patients with different progressions. Clustering patients is particularly important in precision medicine to enhance prediction accuracy for tailored interventions.
 
In the following, we apply the proposed HM model to a real-world scenario using  medical longitudinal data collected at the Anesthesiology and Reanimation Department of the Akdeniz University Hospital as presented and described in \citet{aktas14} (data are freely available at  \url{https://peerj.com/articles/648/}). We first illustrate the data, and then we interpret the results provided by the proposed inferential procedures illustrated in Section \ref{sec:pen}.

\subsection{Data description}
The data include information on patients undergoing spinal anesthesia during a surgery, with  records for $n = 417 $ patients collected from January 2008 to January 2011. Following \citet{aktas14}, we only consider patients aged 17 and over, resulting in $n = 375$ individuals. Various complications can arise during spinal anesthesia, such as hypotension \citep{sharma97, sombo08}, which involves a decrease in mean systolic blood pressure that can potentially lead to death \citep{sanborn96}. The data were collected to monitor the evolution of this phenomenon during surgery to ensure patient safety.
Hypotensive status can be defined according to various criteria. In the study by \citet{aktas14}, which follows one of the most common criteria \citep{klohr10},  the binary response variable is defined such that $y=1$ (hypotensive status) if the systolic blood pressure (SBP) is less than 100 mmHg or less than 80\% of the baseline SBP, and $y=0$ (non-hypotensive status) otherwise.
Measurements occur 8 times, at equally spaced intervals over a period of 40 minutes. 
\begin{figure}[t!] 
    \centering
    \caption{Occurrences of hypotension during spinal anesthesia for each patient; measurements are taken at five-minute intervals, starting from 5 minutes after beginning of surgery}
    \vspace{0.1cm}
    \includegraphics[width=11.5cm,clip=false]{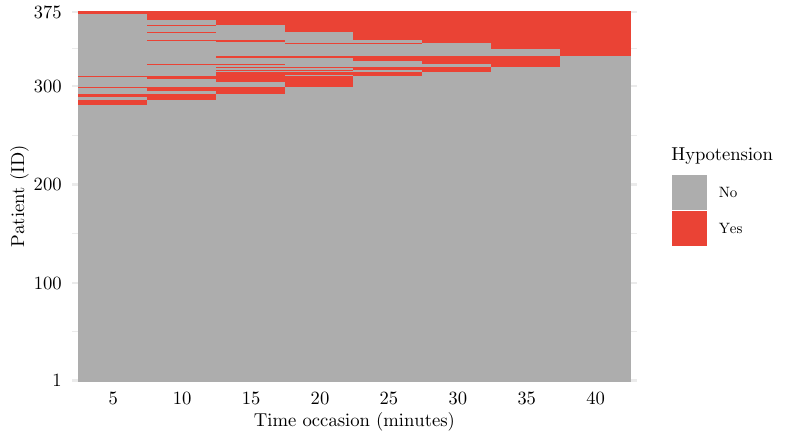}
    \label{fig:response} 
\end{figure}
Individual-specific characteristics  are also recorded, including  time-fixed covariates such as gender, type of surgical hospital unit (general surgery, urology, obstetrics and gynecology), 
position of the patients during the surgery (lithotomy, supine), electrocardiography (ECG) status (normal, abnormal), and doses of medication in the blood (Marcain-heavy, Chirocaine, Fentanyl and Midazolam). 
Time-varying covariates include instead age, diastolic blood pressure (DBP), patient pulse rate.

Figure \ref{fig:response} shows a lasagna plot indicating the presence or absence of hypotension across the 8 measurements for each patient. It reveals that hypotension is a rare event, 
since approximately 25\% ($n=94$) of the patients recorded at least one hypotensive episode during this time.
These findings align with the literature, which reports an incidence of hypotension among 15\%-33\% of patient cases \citep{carpenter92, hartmann02, lin08}.
The average age of patients in this study is approximately 49 years old. Out of the $375$ individuals included, 56\% ($n=210$) were male, and the rest 44\% ($n=165$) female. Surgeries were conducted in three  different hospital units: 38\% in obstetrics and gynecology, 44\% ($n=165$) in urology, and the remaining 18\% ($n=66$) in general surgery. Among the cases, 58\% ($n=218$) of  procedure were performed with patients in the supine position, while 42\% ($n=157$)  were in the lithotomy position. The ECG status was normal for 98\% of patients ($n=366$), while 2\% ($n=9$) had abnormal readings.
\begin{table}[t!]
	\centering
	\caption{Descriptive statistics of anesthetic drugs dose}
	\vspace{-0.1cm}
	\begin{tabular}{lcc@{ $\pm$ }c}
		\toprule
		Anesthetic drug	& Frequency	& Mean	& SD	\\
		\midrule
		Marcain-heavy	& 229		& 7.115 & 6.181 \\
		Midazolam		& 202		& 0.821 & 1.003 \\
		Chirocaine		& 143		& 5.293 & 7.895 \\
		Fentanyl		& 116		& 0.033 & 0.059 \\
		\bottomrule
	\end{tabular}
	\label{tab:stats_anestheticdrugs}
\end{table}
Summary statistics of the doses administered for each drug during anesthesia are presented in Table \ref{tab:stats_anestheticdrugs}.  
Note that each patient may receive more than one medication.

\subsection{Results}
The HM model is specified  by including all the covariates as well as  the lagged hypotension status. This approach measures state dependence and relaxes the  conditional independence assumption. It considers the effect of having hypotension at one  occasion  on the probability of having hypotension at the subsequent occasion, in addition  to the effect of other observable and unobservable factors that may influence this outcome.
\begin{table}[t!]
	\centering
	\caption{Cross-validated log-likelihood ($ \hat{\ell}_{CV} $) for the HM model estimated with number of hidden states ($k$) ranging between 1 and 4, and the penalty parameter ($\lambda$) ranging among 0.00, 0.01, 0.05, 1.00, and 5.00}
	\vspace{-0.1cm}
	\begin{tabular}{lrrrrr}
		\toprule
		\multirow{2}[3]{*}{\centering{$k$}} & \multicolumn{5}{c}{$\lambda$} \\
		\cmidrule(lr){2-6}
			& \multicolumn{1}{c}{$0.00$} & \multicolumn{1}{c}{$0.01$} & \multicolumn{1}{c}{$0.05$} & \multicolumn{1}{c}{$1.00$} & \multicolumn{1}{c}{$5.00$} \\
		\midrule
		$1$	& -53.674	& -			& -			& -			& -			\\
		$2$ & -51.970	& -51.790	& -52.242	& -55.955	& -62.022	\\
		$3$ & -52.968	& -51.153	& -53.316	& -56.585	& -53.907	\\
		$4$	& -53.201	& -51.799	& -52.761	& -58.430	& -56.250	\\
		\bottomrule
	\end{tabular}
	\label{tab:cross-loglikli}
\end{table}
\begin{table}[!ht]
	\centering
	\caption{Estimates of the  regression parameters affecting the conditional response probabilities, along with their standard errors and corresponding  $p$-values under the HM model whit $k=3$ hidden states and $\lambda=0.01$.  Significance is indicated at the $^\dag$10\%, $^*$5\%, and $^{**}$1\% levels}
	\vspace{-0.1cm}
	\begin{tabular}{lr@{\,}lrr}
		\toprule
    	{Covariate}	& \multicolumn{2}{c}{{$\hat{\beta}$}} & \multicolumn{1}{c}{{${s.e.}$}}	& \multicolumn{1}{c}{{$p$-value}} \\
		\midrule
 		Gender (Female) 				&  1.541	& $^*$		& 0.758		& 0.042		\\
  		Position (Supine) 				&  0.813	& 			& 0.509		& 0.110		\\
  		Operation (Urogoly) 			&  0.363	& 			& 1.074		& 0.735		\\
  		Operation (General surgery) 	& -0.061	& 			& 0.895		& 0.946		\\
  		ECG (Normal) 					& -0.878	& 			& 0.897		& 0.328		\\
  		Age (year) 						&  0.037	& $^*$		& 0.019		& 0.052		\\
  		DBP 							& -0.167	& $^{**}$	& 0.022		& 0.000		\\
  		Pulse rate 						& -0.002	& 			& 0.011		& 0.856		\\
  		Marcain-heavy 					& -0.049	& 			& 0.040		& 0.221		\\
 		Midazolam 						&  0.243	& $^*$		& 0.115		& 0.035		\\
  		Chirocaine 						& -0.049	& 			& 0.036		& 0.174		\\ 
  		Fentanyl 						&  1.215	& 			& 2.359		& 0.607		\\
  		Hypotension $(t-1)$ 			&  2.675	& $^{**}$	& 0.260		& 0.000		\\
		\bottomrule
	\end{tabular}
	\label{tab:beestimates}
\end{table}

As illustrated in Section \ref{sec:pen}, we  use the cross-validated log-likelihood to jointly select the number of hidden states and the roughness of the penalty of the HM model. 
The results, as reported in Table \ref{tab:cross-loglikli}, indicate that  $k=3$ hidden states and a penalization parameter of $\lambda = 0.01$ are optimal. This model achieves a 
cross-validated log-likelihood of -51.153 with 23 parameters.

Table \ref{tab:beestimates}  presents 
the estimates of the parameters affecting the conditional distribution of the response variables given the latent process
along with the estimated standard errors and significance levels.
\begin{table}[t!]
	\centering
	\caption{Support points, initial probabilities, and transition probabilities values under the HM model with $k=3$ hidden states and $\lambda=0.01$}
	\vspace{-0.1cm}
	\begin{tabular}{lrrr}
		\toprule
		\multirow{2}[3]{*}{\centering{Estimate}} & \multicolumn{3}{c}{$u$} \\
		\cmidrule(lr){2-4}
							& \multicolumn{1}{c}{${1}$}	& \multicolumn{1}{c}{${2}$}	& \multicolumn{1}{c}{${3}$} \\
 		\midrule
  		${\hat\alpha_{u}}$	& -0.827 					& 3.147 					& 7.359 \\
  		\midrule 
		$\hat\pi_{u}$		& 0.816 					& 0.161 					& 0.023 \\
		\midrule
  		$\hat\pi_{u|1}$ 	& 0.991 					& 0.009 					& 0.000 \\ 
  		$\hat\pi_{u|2}$		& 0.030 					& 0.956 					& 0.014 \\ 
  		$\hat\pi_{u|3}$ 	& 0.114 					& 0.000 					& 0.886 \\  
 		\bottomrule
	\end{tabular}
	\label{tab:latent}
\end{table}
Gender (female) has a significant positive effect on the response variable, indicating that the conditional probability of experimenting hypotension given the hidden state is higher for females. 
Older individuals exibit higher log-odds of being diagnosed with hypotension compared to younger individuals. 
Additionally, the supine position is associated with higher log-odds of hypotension compared to lithotomy position. 
In contrast, DBP has a significant negative effect on the log-odds of hypotension, meaning that lower DBP is associated with higher odds of experiencing hypotension.

Regarding the drugs used, all anesthetics reduce blood pressure and they are thus associated with a hypotensive state. Midazolam has a significant positive effect, indicating that higher concentration of this drug  in the blood is associated with increased odds of experiencing hypotension during surgery. For the other drugs, the estimated coefficients are not significant. 
The lagged response has a significant positive effect on hypotension indicating  serial correlation. 

\begin{figure}[t!]  
    \centering
    \caption{Estimated probability of hypotension under the HM model with $k=3$ hidden states and $\lambda=0.01$ for each state $u$ at every time occasion}
    \vspace{-0.1cm}
    \includegraphics[width=11.5cm,clip=false]{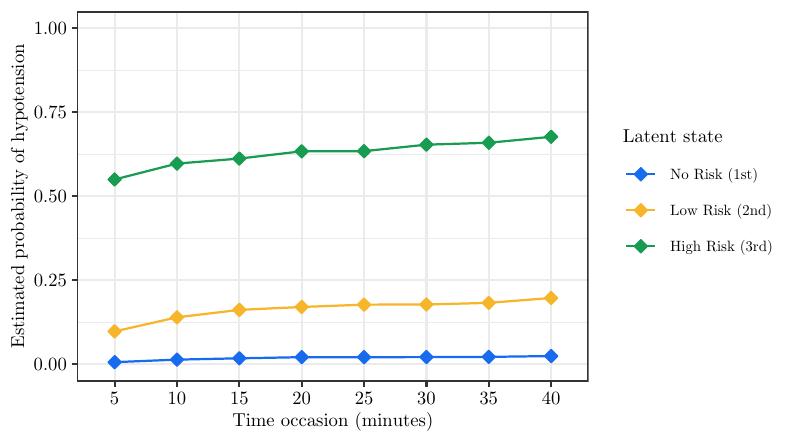}        
    \label{fig:hypo_prob} 
\end{figure}

Table \ref{tab:latent} shows the estimated parameters for the support points of  the hidden states $\alpha_{u}$ as defined in (\ref{eq:logistic}), along with  the initial and transition probabilities. 
The estimated support points provide clinically meaningful labels for each subpopulation of patients identified by the HM model,  indicating that the hidden states can be ordered. The 1st subpopulation corresponds to patients with the lowest propensity for hypotension, while the 3rd corresponds to those with the highest propensity.
Using equation (\ref{eq:logistic}), we  estimate the conditional probability of hypothension for each subpopulation over time, as shown in Figure \ref{fig:hypo_prob}. Patients in the 1st subpopulation have an almost negligible probability of hypotension throughout  the surgery. Patients  in the 2nd subpopulation experience a low probability of hypotension, ranging approximately from 0.10 to 0.20. In contrast, patients in the 3rd subpopulation have a very high probability of hypotension during the surgery, ranging from 0.54 to 0.68.
Additionally, for individuals in the 3rd subpopulation the probability  of hypotension increases  over time. Therefore, the groups can be labeled as follows: the 1st subpopulation  as ``No Risk'' the 2nd as ``Low Risk'' and the 3rd as ``High Risk''.

The estimated initial probabilities reported in Table \ref{tab:latent} suggest that,  at the first time occasion, approximately $82\%$ of patients belong to the 1st subpopulation  (No Risk), $16\%$ to the 2nd subpopulation  (Low Risk), and 2\% to the 3rd subpolulation (High and Increasing Risk). The estimated transition probabilities suggest high persistence within the 1st and 2nd subpopulations, with about 1.4\% of patients transitioning from the low-risk (2nd) to the high-risk (3rd) subgroup. In contrast, the high-risk subgroup shows less persistence, with 11.4\% of patients estimated to transition to the no-risk group.

Patients are assigned to the hidden states using local decoding as described in Section \ref{sec:model} based on the estimated posterior probabilities. Each patient at each time occasion is assigned to the subgroup with the highest posterior membership probability.
According to this estimated posterior distribution, approximately 86.7-87.6\% of the individuals are classified in the 1st subgroup (No Risk) over the considered time period, 10.9-11.5\% in the 2nd subgroup (Low Risk), and 1.6-2.1\% in the 3rd subgroup (High Risk). 
The model can assist doctors in identifying patients at high risk of hypotension during the initial moments of surgery, enabling them to take timely preventive measures to ensure patient safety.

\begin{figure}[ht]  
	\centering
	\caption{Estimated average individual latent effect  ($\hat{\bar{\alpha}}$) at each time occasion for 7 individuals classified at the last period as follows: the left panel ($a$) refers to patients classified as No Risk (1st state), middle panel ($b$) refers those classified as Low Risk (2nd state), and right panel ($c$) refers to those classified as High Risk (3rd state)}
	\vspace{-0.1cm}
	\includegraphics[width=12.5cm,clip=false]{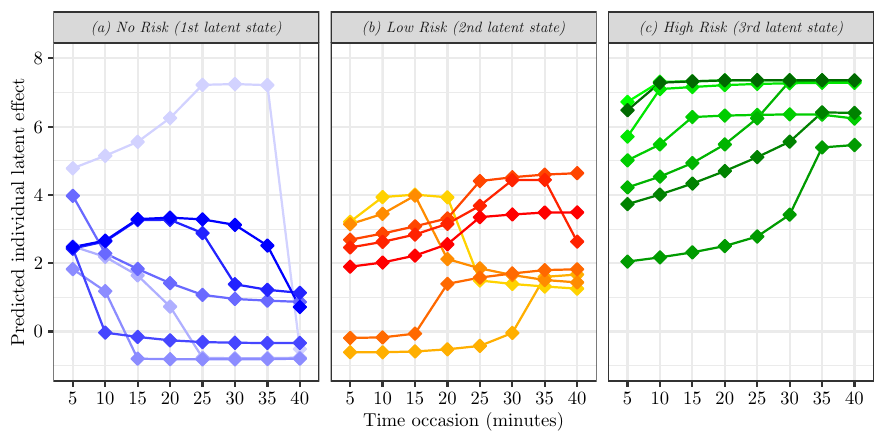}
	\label{fig:alphas} 
\end{figure}

Figure \ref{fig:alphas} shows the values $\hat{\bar{\alpha}}_i^{(t)}$ estimated for selected individuals. 
Specifically, the left panel ($a$)  refers to 7 patients clustered in the 1st subgroup (No Risk) at the last measurement, which show a general decreasing trend in the estimated values over time, indicating a reducing risk of hypotension for these patients.
The middle panel ($b$) concerns other 7 patients clustered in the 2nd subgroup (Low Risk) at the time occasion. Their trajectories are more heterogeneous over time, with some patients experiencing a slight increase and others a slight decrease in the risk during surgery. 
The right panel ($c$)  relates to 7 patients clustered in the 3rd subgroup (High Risk) at the last time occasion, indicating that these patients tend to have an increasing risk of hypotension during the observed period.

It is worth mentioning that the HM model proposed here differs substantially from both the generalized estimating equation method and the generalized linear mixed model employed to analyze the same data in \cite{aktas14}. The first approach describes changes in the population mean, while the second approach analyzes changes in individual response means. 
These models employ almost the same number of parameters of our proposal; however, for both, it is necessary to estimate a breakpoint using the 20-minute time point to better account for time dynamics, as such time effects are significant in each model. 
The estimated effects are higher in magnitude with the HM model. Unlike the other two models, gender, position, and  Midazolam have a significant effect in the HM model. It is important to note that some drugs are still the subject of research concerning their role in inducing hypotension \citep{pali:24}. 
The HM model accounts for marginal effects, and within the proposed formulation, individual heterogeneity is captured by the random cluster effects. This makes it possible, for example, to disentangle the stabilization effect seen in Figure \ref{fig:alphas}, which is instead parameterized through the breakpoint in the other two  models presented in the original article analyzing these data.

\section{Conclusions}\label{sec:con}
In this study, we introduce a new estimation method for hidden Markov models where covariates directly affect the response variables. The model accounts for the so-called  unobserved heterogeneity and serial dependence by including the lagged response among the covariates. We propose a penalized maximum likelihood estimation method to prevent instability in the estimates, through a penalization term aimed at reducing the separation of latent states. Additionally, we introduce a cross-validation approach to determine both the necessity of a penalty and the appropriate number of latent states. This method serves as an alternative to traditional information criteria for selecting the optimal model when the number of unknown subpopulations is not known a priori. The HM model is particularly useful for dynamically classifying subjects into groups based on the estimated posterior probabilities, and for detecting individual trajectories and their evolution over time.

The proposal is validated through an extensive simulation study, where we assess the effectiveness of the procedure under different scenarios. We demonstrate improved accuracy in parameter estimates while reducing computational time. The empirical illustration shows that the model can detect the effects of significant covariates such as gender and the type of drug used during surgery for hypotension. It is also effective in classifying patients into groups with increasing severity, making it particularly valuable for real-time monitoring of high-risk patients during surgery, thereby acting as an early warning system. Additionally, we show that the proposed model differs from alternative models commonly used for the analysis of longitudinal data, particularly because it allows for straightforward accounting of serial dependence and identification of patient clusters with varying hypotension complexities and trajectories over time.

We note that the proposed penalty term corresponds to the one used in Ridge regression. Consequently, the estimates of the support points obtained with our Ridge-type penalized approach are equivalent to the ones provided in a Bayesian framework when Gaussian priors are assumed for the $\alpha_u$; see \citet{tibs96} for more details. 
Furthermore, an alternative approach could involve a Lasso-type penalization, which corresponds to the use of Laplace priors in the Bayesian estimation.

Future research is needed to investigate the circumstances under which large estimates of the support points of latent states occur. Additionally, efforts should be directed towards developing a faster estimation procedure, as the current computational time for estimating the HM model with covariates affecting the response, whether with or without penalization, is high.
This issue can be addressed by implementing the estimation procedure in C++, thus also reducing the time required for cross-validation. Furthermore, the model can be extended to accommodate multivariate response variables of mixed types, including both continuous and categorical variables, as well as to handle missing values. 
The proposal can also be enhanced to account for lagged dependencies of second or higher order. Finally, methods to achieve scalability, such as parallel computation and dimensitionality reduction, will be essential for efficiently handling large datasets.

\paragraph{Acknowledgment.} The authors acknowledge the financial support from the grant ``Hidden Markov Models for Early Warning Systems'' of Ministero dell’Universit\`a e della Ricerca (PRIN 2022TZEXKF) funded by the European Union - Next Generation EU, Mission 4, Component 2, CUP J53D23004990006.

\appendix
\section{Appendix}\label{app}
Referring to the simulated scenarios introduced in Section \ref{sec:sim} we show the  percentage variation in the MSE values between true and estimated parameter for the covariate regression coefficients $\beta_1$, $\beta_2$, $\beta_3$, and $\beta_4$ (Tables \ref{tab:beta1}, \ref{tab:beta2}, \ref{tab:beta3}, and \ref{tab:beta4}, respectively). 
\begin{table}[h!]
	\centering
	\caption{Percentage variation in the mean squared error between the true and the estimated value of parameter $\beta_1$ for the penalized estimation ($\lambda = 0.01$ and $\lambda = 0.05$) with respect to the standard maximum likelihood estimation ($\lambda = 0.00$)}
	\scriptsize
	\begin{tabular}{llllrrrrr}
		\toprule
		& & & & \multicolumn{1}{c}{$\bm{\alpha}^1$}	& \multicolumn{1}{c}{$\bm{\alpha}^2$}	& \multicolumn{1}{c}{$\bm{\alpha}^3$}	& \multicolumn{1}{c}{$\bm{\alpha}^4$}	& \multicolumn{1}{c}{$\bm{\alpha}^5$} \\
		\cmidrule(lr){5-9}
		$n$	& $T$	& $\bm{\Pi}$	& $\lambda$	& \multicolumn{5}{c}{Percentage variation (\%)}	\\
		\midrule
		\multirow{8}[5]{*}{$250$}	& \multirow{4}[2]{*}{$10$}	& \multirow{2}{*}{High}	& $0.01$	& $-$8.50	& $-$11.11	& $-$12.33	& $-$99.44	& $-$99.95	\\
									& 							& 						& $0.05$	& $-$27.71	& $-$19.60	& $-$41.17	& $-$99.76	& $-$99.97	\\
		\cmidrule(lr){3-9}
									& 							& \multirow{2}{*}{Low}	& $0.01$	& $-$8.36	& $-$36.00	& $-$18.62	& $-$99.97	& $-$99.87	\\
									& 							& 						& $0.05$	& $-$30.93	& $-$66.67	& $-$56.95	& $-$99.98	& $-$99.90	\\
		\cmidrule(lr){2-9}
									& \multirow{4}[2]{*}{$20$}	& \multirow{2}{*}{High}	& $0.01$	& $-$1.08	& $-$2.28	& $-$0.31	& $-$61.31	& $-$99.89	\\
									& 							& 						& $0.05$	& $-$7.83	& $-$5.68	& $-$7.79	& $-$84.88	& $-$99.92	\\
		\cmidrule(lr){3-9}
									& 							& \multirow{2}{*}{Low}	& $0.01$	& $-$2.32	& $-$18.08	& $-$4.26	& $-$99.54	& $-$98.54	\\
									& 							& 						& $0.05$	& $-$7.27	& $-$33.79	& $-$14.13	& $-$99.82	& $-$98.83	\\
		\midrule
		\multirow{8}[5]{*}{$500$}	& \multirow{4}[2]{*}{$10$}	& \multirow{2}{*}{High}	& $0.01$	& $-$1.45	& $-$6.94	& $+$0.23	& $-$61.41	& $-$99.62	\\
									& 							& 						& $0.05$	& $-$9.36	& $-$9.31	& $-$6.48	& $-$87.89	& $-$99.65	\\
		\cmidrule(lr){3-9}
									& 							& \multirow{2}{*}{Low}	& $0.01$	& $-$1.02	& $-$22.13	& $-$16.41	& $-$99.94	& $-$99.60	\\
									& 							& 						& $0.05$	& $-$17.08	& $-$43.58	& $-$30.40	& $-$99.96	& $-$99.66	\\
		\cmidrule(lr){2-9}
									& \multirow{4}[2]{*}{$20$}	& \multirow{2}{*}{High}	& $0.01$	& $-$1.62	& $+$1.10	& $-$2.66	& $-$17.92	& $-$99.15	\\
									& 							& 						& $0.05$	& $-$4.82	& $+$3.33	& $-$6.00	& $-$48.82	& $-$99.28	\\
		\cmidrule(lr){3-9}
									& 							& \multirow{2}{*}{Low}	& $0.01$	& $-$2.42	& $-$1.17	& $-$2.25	& $-$38.90	& $-$86.17	\\
									& 							& 						& $0.05$	& $-$0.16	& $+$9.65	& $-$13.44	& $-$51.46	& $-$87.41	\\
		\bottomrule
	\end{tabular}
	\label{tab:beta1}
\end{table}

\begin{table}[t!]
	\centering
	\caption{Percentage variation in the mean squared error between the true and the estimated value of parameter $\beta_2$ for the penalized estimation ($\lambda = 0.01$ and $\lambda = 0.05$) with respect to the standard maximum likelihood estimation ($\lambda = 0.00$)}
	\scriptsize
	\begin{tabular}{llllrrrrr}
		\toprule
		& & & & \multicolumn{1}{c}{$\bm{\alpha}^1$}	& \multicolumn{1}{c}{$\bm{\alpha}^2$}	& \multicolumn{1}{c}{$\bm{\alpha}^3$}	& \multicolumn{1}{c}{$\bm{\alpha}^4$}	& \multicolumn{1}{c}{$\bm{\alpha}^5$} \\
		\cmidrule(lr){5-9}
		$n$	& $T$	& $\bm{\Pi}$	& $\lambda$	& \multicolumn{5}{c}{Percentage variation (\%)}	\\
		\midrule
		\multirow{8}[5]{*}{$250$}	& \multirow{4}[2]{*}{$10$}	& \multirow{2}{*}{High}	& $0.01$	& $-$6.40	& $-$11.58	& $-$9.28	& $-$99.94	& $-$99.93	\\
									& 							& 						& $0.05$	& $-$23.99	& $-$30.67	& $-$35.59	& $-$99.97	& $-$99.96	\\
		\cmidrule(lr){3-9}
									& 							& \multirow{2}{*}{Low}	& $0.01$	& $-$7.68	& $-$37.81	& $-$18.43	& $-$99.95	& $-$99.92	\\
									& 							& 						& $0.05$	& $-$24.85	& $-$64.86	& $-$53.78	& $-$99.96	& $-$99.92	\\
		\cmidrule(lr){2-9}
									& \multirow{4}[2]{*}{$20$}	& \multirow{2}{*}{High}	& $0.01$	& $+$1.14	& $-$1.62	& $+$0.99	& $-$60.18	& $-$99.87	\\
									& 							& 						& $0.05$	& $-$5.36	& $-$13.31	& $-$4.40	& $-$82.53	& $-$99.93	\\
		\cmidrule(lr){3-9}
									& 							& \multirow{2}{*}{Low}	& $0.01$	& $-$3.91	& $-$14.31	& $-$4.42	& $-$99.76	& $-$99.55	\\
									& 							& 						& $0.05$	& $-$7.10	& $-$18.82	& $-$27.18	& $-$99.90	& $-$99.62	\\
		\midrule
		\multirow{8}[5]{*}{$500$}	& \multirow{4}[2]{*}{$10$}	& \multirow{2}{*}{High}	& $0.01$	& $-$1.26	& $-$9.56	& $+$0.07	& $-$45.09	& $-$99.79	\\
									& 							& 						& $0.05$	& $-$11.99	& $-$23.08	& $-$10.39	& $-$65.75	& $-$99.83	\\
		\cmidrule(lr){3-9}
									& 							& \multirow{2}{*}{Low}	& $0.01$	& $+$0.95	& $-$26.14	& $-$16.43	& $-$99.94	& $-$99.67	\\
									& 							& 						& $0.05$	& $-$24.57	& $-$56.25	& $-$30.24	& $-$99.97	& $-$99.71	\\
		\cmidrule(lr){2-9}
									& \multirow{4}[2]{*}{$20$}	& \multirow{2}{*}{High}	& $0.01$	& $-$3.42	& $+$0.50	& $-$2.42	& $-$9.16	& $-$99.36	\\
									& 							& 						& $0.05$	& $-$5.54	& $+$3.79	& $-$3.89	& $-$15.73	& $-$99.48	\\
		\cmidrule(lr){3-9}
									& 							& \multirow{2}{*}{Low}	& $0.01$	& $-$1.17	& $-$4.66	& $+$0.08	& $-$22.69	& $-$95.54	\\
									& 							& 						& $0.05$	& $+$0.90	& $-$2.54	& $-$7.29	& $-$12.89	& $-$96.29	\\
		\bottomrule
	\end{tabular}
	\label{tab:beta2}
\end{table}

\begin{table}[t!]
\centering
\caption{Percentage variation in the mean squared error between the true and the estimated value of parameter $\beta_3$ for the penalized estimation ($\lambda = 0.01$ and $\lambda = 0.05$) with respect to the standard maximum likelihood estimation ($\lambda = 0.00$)}
\scriptsize
\begin{tabular}{llllrrrrr}
	\toprule
	& & & & \multicolumn{1}{c}{$\bm{\alpha}^1$}	& \multicolumn{1}{c}{$\bm{\alpha}^2$}	& \multicolumn{1}{c}{$\bm{\alpha}^3$}	& \multicolumn{1}{c}{$\bm{\alpha}^4$}	& \multicolumn{1}{c}{$\bm{\alpha}^5$} \\
	\cmidrule(lr){5-9}
	$n$	& $T$	& $\bm{\Pi}$	& $\lambda$	& \multicolumn{5}{c}{Percentage variation (\%)}	\\
	\midrule
	\multirow{8}[5]{*}{$250$}	& \multirow{4}[2]{*}{$10$}	& \multirow{2}{*}{High}	& $0.01$	& $-$6.62	& $-$12.28	& $-$11.79	& $-$99.94	& $-$99.96	\\
								& 							& 						& $0.05$	& $-$26.14	& $-$25.00	& $-$34.05	& $-$99.98	& $-$99.97	\\
	\cmidrule(lr){3-9}
								& 							& \multirow{2}{*}{Low}	& $0.01$	& $-$8.05	& $-$35.50	& $-$20.62	& $-$99.98	& $-$99.93	\\
								& 							& 						& $0.05$	& $-$32.43	& $-$62.08	& $-$57.09	& $-$99.98	& $-$99.95	\\
	\cmidrule(lr){2-9}
								& \multirow{4}[2]{*}{$20$}	& \multirow{2}{*}{High}	& $0.01$	& $-$1.02	& $-$3.47	& $+$0.09	& $-$75.87	& $-$99.89	\\
								& 							& 						& $0.05$	& $-$2.28	& $-$11.78	& $-$8.39	& $-$90.58	& $-$99.94	\\
	\cmidrule(lr){3-9}
								& 							& \multirow{2}{*}{Low}	& $0.01$	& $-$1.91	& $-$19.77	& $+$2.99	& $-$99.79	& $-$99.03	\\
								& 							& 						& $0.05$	& $-$5.46	& $-$33.61	& $-$22.83	& $-$99.88	& $-$99.21	\\
	\midrule
	\multirow{8}[5]{*}{$500$}	& \multirow{4}[2]{*}{$10$}	& \multirow{2}{*}{High}	& $0.01$	& $-$0.58	& $-$1.39	& $-$2.12	& $-$56.70	& $-$99.76	\\
								& 							& 						& $0.05$	& $-$9.96	& $+$0.94	& $-$10.04	& $-$79.08	& $-$99.83	\\
	\cmidrule(lr){3-9}
								& 							& \multirow{2}{*}{Low}	& $0.01$	& $+$0.23	& $-$25.51	& $-$10.25	& $-$99.96	& $-$99.37	\\
								& 							& 						& $0.05$	& $-$22.8	& $-$54.45	& $-$27.71	& $-$99.95	& $-$99.50	\\
	\cmidrule(lr){2-9}
								& \multirow{4}[2]{*}{$20$}	& \multirow{2}{*}{High}	& $0.01$	& $-$2.35	& $-$0.63	& $-$3.16	& $-$19.69	& $-$99.41	\\
								& 							& 						& $0.05$	& $-$3.55	& $-$2.06	& $-$5.73	& $-$34.37	& $-$99.54	\\
	\cmidrule(lr){3-9}
								& 							& \multirow{2}{*}{Low}	& $0.01$	& $-$2.49	& $-$3.44	& $-$6.86	& $-$35.29	& $-$94.24	\\
								& 							& 						& $0.05$	& $-$2.27	& $+$2.26	& $-$21.08	& $-$46.01	& $-$95.22	\\
	\bottomrule
\end{tabular}
\label{tab:beta3}
\end{table}

\begin{table}[t!]
\centering
\caption{Percentage variation in the mean squared error between the true and the estimated value of parameter $\beta_4$ for the penalized estimation ($\lambda = 0.01$ and $\lambda = 0.05$) with respect to the standard maximum likelihood estimation ($\lambda = 0.00$)}
\scriptsize
\begin{tabular}{llllrrrrr}
	\toprule
	& & & & \multicolumn{1}{c}{$\bm{\alpha}^1$}	& \multicolumn{1}{c}{$\bm{\alpha}^2$}	& \multicolumn{1}{c}{$\bm{\alpha}^3$}	& \multicolumn{1}{c}{$\bm{\alpha}^4$}	& \multicolumn{1}{c}{$\bm{\alpha}^5$} \\
	\cmidrule(lr){5-9}
	$n$	& $T$	& $\bm{\Pi}$	& $\lambda$	& \multicolumn{5}{c}{Percentage variation (\%)}	\\
	\midrule
	\multirow{8}[5]{*}{$250$}	& \multirow{4}[2]{*}{$10$}	& \multirow{2}{*}{High}	& $0.01$	& $-$2.07	& $-$3.09	& $-$1.46	& $-$99.31	& $-$99.97	\\
								& 							& 						& $0.05$	& $-$5.67	& $-$12.66	& $-$1.90	& $-$99.66	& $-$99.99	\\
	\cmidrule(lr){3-9}
								& 							& \multirow{2}{*}{Low}	& $0.01$	& $-$0.00	& $-$19.21	& $+$5.05	& $-$99.95	& $-$99.97	\\
								& 							& 						& $0.05$	& $+$1.17	& $-$45.83	& $-$9.28	& $-$99.98	& $-$99.98	\\
	\cmidrule(lr){2-9}
								& \multirow{4}[2]{*}{$20$}	& \multirow{2}{*}{High}	& $0.01$	& $+$1.21	& $-$1.64	& $-$0.09	& $-$62.73	& $-$99.93	\\
								& 							& 						& $0.05$	& $-$1.01	& $-$3.77	& $-$2.40	& $-$86.33	& $-$99.96	\\
	\cmidrule(lr){3-9}
								& 							& \multirow{2}{*}{Low}	& $0.01$	& $+$0.52	& $-$5.98	& $-$2.71	& $-$99.12	& $-$99.66	\\
								& 							& 						& $0.05$	& $+$1.61	& $-$14.67	& $-$44.16	& $-$99.71	& $-$99.81	\\
	\midrule
	\multirow{8}[5]{*}{$500$}	& \multirow{4}[2]{*}{$10$}	& \multirow{2}{*}{High}	& $0.01$	& $+$0.94	& $+$0.30	& $-$0.13	& $-$58.92	& $-$99.89	\\
								& 							& 						& $0.05$	& $+$1.04	& $+$1.46	& $-$0.66	& $-$84.81	& $-$99.92	\\
	\cmidrule(lr){3-9}
								& 							& \multirow{2}{*}{Low}	& $0.01$	& $+$0.11	& $-$12.60	& $-$8.10	& $-$99.91	& $-$99.77	\\
								& 							& 						& $0.05$	& $-$8.21	& $-$36.85	& $-$11.68	& $-$99.98	& $-$99.88	\\
	\cmidrule(lr){2-9}
								& \multirow{4}[2]{*}{$20$}	& \multirow{2}{*}{High}	& $0.01$	& $-$1.13	& $+$0.61	& $-$1.13	& $-$22.65	& $-$99.64	\\
								& 							& 						& $0.05$	& $-$0.75	& $+$2.10	& $-$0.58	& $-$51.71	& $-$99.75	\\
	\cmidrule(lr){3-9}
								& 							& \multirow{2}{*}{Low}	& $0.01$	& $-$1.93	& $-$1.67	& $-$0.19	& $-$19.07	& $-$97.1	\\
								& 							& 						& $0.05$	& $-$0.56	& $-$8.34	& $+$9.12	& $-$25.37	& $-$98.03	\\
	\bottomrule
\end{tabular}
\label{tab:beta4}
\end{table}

\newpage
\addcontentsline{toc}{section}{References}
\bibliographystyle{chicago}
\bibliography{BiblioPenHM}

\end{document}